# Quantum and Classical Disparity and Accord

**Mario Rabinowitz**

**Abstract** Discrepancies and accords between quantum (QM) and classical mechanics (CM) related to expectation values and periods are found for both the simple harmonic oscillator (SHO) and a free particle in a box (FPB), which may apply generally. These indicate non-locality is expected throughout QM. The FPB energy states violate the Correspondence Principle. Previously unexpected accords are found and proven that $\langle x^2 \rangle_{CM} = \langle x^2 \rangle_{QM}$ and $t_{CM} = t_{QMb}$ (beat period, i.e. beats between the phases for adjoining energy states) for the SHO for all quantum numbers, *n*. However, for the FPB the beat periods differ significantly at small *n*. It is shown that a particle's velocity in an infinite square well varies, no matter how wide the box, nor how far the particle is from the walls. The quantum free particle variances share an indirect commonality with the Aharonov-Bohm and Aharonov-Casher effects in that there is a quantum action in the absence of a force. The concept of an "Expectation Value over a Partial Well Width" is introduced. This paper raises the question as to whether these inconsistencies are undetectable, or can be empirically ascertained. These inherent variances may need to be fixed, or nature is manifestly more non-classical than expected.

## 1 Introduction

Although this paper focuses on quantum mechanical (QM) and classical mechanical (CM) discrepancies, noteworthy consonances, for all quantum numbers, were found that $\langle x^2 \rangle_{CM} = \langle x^2 \rangle_{QM}$ and $t_{QMb} = t_{CM}$ (beat periods) for the harmonic oscillator. This result is unique because by the Virial Theorem the harmonic oscillator is the only case where $\langle PE \rangle = \langle KE \rangle = \frac{1}{2} E_{total} = \frac{1}{2} k \langle x^2 \rangle$ in both CM and QM making $\langle x^2 \rangle_{QM} = \langle x^2 \rangle_{CM}$ as proven in detail Sec. 3.3.

Mario Rabinowitz
Armor Research; 715 Lakemead Way; Redwood City, CA 94062-3922 USA
e-mail Mario715@gmail.com

However, other quantum and classical discrepancies persist even at large quantum numbers. This is a violation of the Correspondence Principle, and indicates that QM may not be a theory that applies in all cases of the realm of observation. These and other disparities are analyzed here, and appear to be both prevalent for all potentials, and possibly testable experimentally. A free particle in a box manifests similarities with the Aharonov-Bohm [1] and Aharonov-Casher [2] effects in that there is a quantum action in the absence of a force. Therefore these established effects will be discussed quantum mechanically and classically to facilitate comparison with the variances found in this paper.

## 2 Partial Well Width Expectation Values for an Infinite Square Well

### 2.1 General Quantum Considerations

Rather than calculate expectation values over the full range in which a particle can be found, in this Section it will be informative to calculate partial well width expectation values (sub-ensembles) to give insight to measurements that are confined to sub-regions of a larger domain. We can find these partial width expectation values, starting with the definition of the expectation value of a variable $a$ in a region e.g. a potential well of width -*a* to *a*.

$$\begin{aligned}\left\langle a_{QM} \right\rangle_{-a,a} &= \int_{-a}^{a} y^* a\, y\, dx = \int_{-a}^{-a/2} y^* a\, y\, dx + \int_{-a/2}^{0} y^* a\, y\, dx + \int_{0}^{a/2} y^* a\, y\, dx + \int_{a/2}^{a} y^* a\, y\, dx \\ &= \langle a \rangle_{-a,-a/2} + \langle a \rangle_{-a/2,0} + \langle a \rangle_{0,a/2} + \langle a \rangle_{a/2,a}\end{aligned} \quad , (2.1)$$

where for clarity and convenience the range -*a* to *a* has been broken up into 4 smaller equal regions, defining each partial expectation value.

Similarly for normalization of the wave function $y$ we have

$$1 = \int_{-a}^{a} y^* y\, dx = \int_{-a}^{-a/2} y^* y\, dx + \int_{-a/2}^{0} y^* y\, dx + \int_{0}^{a/2} y^* y\, dx + \int_{a/2}^{a} y^* y\, dx . \quad (2.2a)$$

The range could have been broken up into any number of different sized regions. The treatment here is one-dimensional for simplicity, but can easily be generalized to any number of dimensions.

The established convention to normalize over the entire range will be followed here. Anomalies can be circumvented for a parameter that is constant in a given state such as energy E in an infinite square well. The partial range normalization from *a1* to *a2* yields

$$\left\langle E_{QM} \right\rangle = \frac{\int_{a1}^{a2} y^* E y dx}{\int_{a1}^{a2} y^* y dx} = \frac{\int_{a1}^{a2} y^* \left[ \frac{-\hbar^2}{2m} \nabla^2 \right] y dx}{\int_{a1}^{a2} y^* y dx} = \frac{E \int_{a1}^{a2} y^* y dx}{\int_{a1}^{a2} y^* y dx} = E. \qquad (2.2b)$$

The normalization in eq. (2.2a) for partial energy expectation values is the sum over partial intervals. The normalization in eq. (2.2b) is for the total energy expectation value over a single partial interval.

Sec. 2 will illustrate that not only can normalization affect the outcome, but also the partitioning of the expectation value intervals. Furthermore in Sec. 2 one may consider that the energy is constant in each interval, but the population density varies. However, Secs. 3 and 4 do not have this option.

**2.2 Quantum Case for Particle in an Infinite Square Well of Width -*a* to *a***

For an Infinite Square Well of Width -*a* to *a*, the full range normalized wave function that satisfies the time-independent part of the Schrödinger non-relativistic wave equation (4.1) is

$$y_n(x) = \sqrt{\tfrac{1}{a}} \sin\left(\tfrac{npx}{2a} - \tfrac{np}{2}\right) \qquad (2.3)$$

The energy expectation value is equal to the Hamiltonian expectation value. For the full well width:

$$\left\langle E_{QM} \right\rangle_{-a,a} = \left\langle H \right\rangle = \left\langle \frac{p^2}{2m} \right\rangle = \int_{-a}^{a} y_n^* \left[ \frac{-\hbar^2}{2m} \nabla^2 \right] y_n dx = \int_{-a}^{a} y_n^* \left[ \frac{-\hbar^2}{2m} \frac{¶^2}{¶x^2} \right] y_n dx = \frac{h^2 n^2}{32 m a^2}. \qquad (2.4)$$

Now for four equal partial well widths, the partial energy expectation values are:

$$\left\langle E_{QM} \right\rangle_{-a,-a/2} = \int_{-a}^{-a/2} y_n^* \left[ \frac{-\hbar^2}{2m} \frac{¶^2}{¶x^2} \right] y_n dx = \frac{h^2 n^2 \left[ np + 2 \sin(3np/2) \right]}{128 m a^2 p}. \qquad (2.5)$$

$$\left\langle E_{QM} \right\rangle_{-a/2,0} = \frac{h^2 n^2 \left[ np - 2\sin\left(3np/2\right)\right]}{128 ma^2 p}. \tag{2.6}$$

$$\left\langle E_{QM} \right\rangle_{0,a/2} = \frac{h^2 n^2 \left[ np + 2\sin\left(np/2\right)\right]}{128 ma^2 p}. \tag{2.7}$$

$$\left\langle E_{QM} \right\rangle_{a/2,a} = \frac{h^2 n^2 \left[ np - 2\sin\left(np/2\right)\right]}{128 ma^2 p}. \tag{2.8}$$

From eqs. (2.5) through (2.8) we see explicitly:

$$\left\langle E_{QM} \right\rangle_{-a,a} = \left\langle E_{QM} \right\rangle_{-a,-a/2} + \left\langle E_{QM} \right\rangle_{-a/2,0} + \left\langle E_{QM} \right\rangle_{0,a/2} + \left\langle E_{QM} \right\rangle_{a/2/,a}. \tag{2.9}$$

By inspection of eqs. (2.5) through (2.8)

$$\left\langle E_{QM} \right\rangle_{-a,-a/2} = \left\langle E_{QM} \right\rangle_{a/2,a} \text{, and} \tag{2.10}$$

$$\left\langle E_{QM} \right\rangle_{-a/2,0} = \left\langle E_{QM} \right\rangle_{0,a/2}. \tag{2.11}$$

Interestingly,

$$\left\langle E_{QM} \right\rangle_{-a,-a/2} = \left\langle E_{QM} \right\rangle_{a/2,a} = \frac{h^2 (p-2)}{128 ma^2 p} \; for\, n = 1\text{, and} \tag{2.12}$$

$$\left\langle E_{QM} \right\rangle_{-a/2,0} = \left\langle E_{QM} \right\rangle_{0,a/2} = \frac{h^2 (p+2)}{128 ma^2 p} \; for\, n = 1. \tag{2.13}$$

Therefore in a force-free region, without the action of a force, although the particle's energy averages out to $E_{total}$ and is conserved for the region as a whole, the particle's local partial energy appears to increase and decrease as the particle goes from sub-region to sub-region, for full range normalization. This is as if there is a non-local quantum mechanical action (as previously discussed by Rabinowitz [17] for $\left\langle x^2 \right\rangle_{-a,a}$, and will be further analyzed in this paper). This is the case for all odd $n$ states.

But equally interesting this does not occur in these particular regions for even $n$ states (odd $y$).

$$\left\langle E_{QM} \right\rangle_{-a,-a/2} = \left\langle E_{QM} \right\rangle_{-a/2,0} = \left\langle E_{QM} \right\rangle_{0,a/2} = \left\langle E_{QM} \right\rangle_{a/2/,a} = \frac{1}{4} \left\langle E_{QM} \right\rangle_{-a,a} \; for\ all\ even\, n\, states. \tag{2.14}$$

In particular:

$$\left\langle E_{QM} \right\rangle_{-a,-a/2} = \left\langle E_{QM} \right\rangle_{-a/2,0} = \left\langle E_{QM} \right\rangle_{0,a/2} = \left\langle E_{QM} \right\rangle_{a/2/,a} = \frac{h^2}{8ma^2} \; for\, n = 2. \tag{2.15}$$

The variations for odd n, and constancy for even n, follow from the properties of $|y|^2$ since $\langle E_{QM} \rangle = \int_{a1}^{a2} E_{total} y^* y\, dx = \int_{a1}^{a2} E_t |y|^2\, dx = E_t \int_{a1}^{a2} |y|^2 dx$.

### 2.3 Classical Case for Particle in an Infinite Square Well of Width -*a* to *a*

We can find classical partial width expectation values similarly to the quantum case, starting with the standard expectation value of a variable $a$ for a particle that is confined to a region e.g. a potential well of width -*a* = x = *a*.

$$\begin{aligned}\langle a_{CM} \rangle_{-a,a} &= \int_{-a}^{a} a bPdx = \int_{-a}^{-a/2} abPdx + \int_{-a/2}^{0} abPdx + \int_{0}^{a/2} abPdx + \int_{a/2}^{a} abPdx \\ &= \langle a \rangle_{-a,-a/2} + \langle a \rangle_{-a/2,0} + \langle a \rangle_{0,a/2} + \langle a \rangle_{a/2,0}\end{aligned} \tag{2.16}$$

where *P* is the classical probability, which is inversely proportional to the particle's velocity, and *b* is the normalization coefficient.

For a classical free particle in a box, *P* is uniform because the particle's speed is constant in the infinite well of width -a = x = *a*. Normalizing the classical probability,

$$1 = \int_{-a}^{a} bPdx = bP(2a) \Rightarrow bP = \frac{1}{2a}. \tag{2.17}$$

The free particle's energy expectation value for the full well width is

$$\langle E_{CM} \rangle_{-a,a} = \int_{-a}^{a} EbPdx = \int_{-a}^{a} E[1/2a]dx = E \tag{2.18}$$

The partial energy expectation values for partial well widths are

$$\langle E_{CM} \rangle_{-a,a/2} = \int_{-a}^{-a/2} EbPdx = \int_{-a}^{-a/2} E[1/2a]dx = E/4 \tag{2.19}$$

$$\langle E_{CM} \rangle_{-a/2,0} = \int_{-a/2}^{0} E[1/2a]dx = E/4 \tag{2.20}$$

$$\langle E_{CM} \rangle_{0,a/2} = \int_{0}^{a/2} E[1/2a]dx = E/4 \tag{2.21}$$

$$\langle E_{CM} \rangle_{a/2,a} = \int_{a/2}^{0} E[1/2a]dx = E/4 \tag{2.22}$$

From eqs. (2.18) through (2.22) we have explicitly:

$$\langle E_{CM} \rangle_{-a,a} = \langle E_{CM} \rangle_{-a,-a/2} + \langle E_{CM} \rangle_{-a/2,0} + \langle E_{CM} \rangle_{0,a/2} + \langle E_{CM} \rangle_{a/2/a} = E. \tag{2.23}$$

Classically the particle has a partial well width energy that is constant across the entire well width. Here the partial energy in each sub-region is $E/4$ because there were 4 sub-regions.

## 3 Simple Harmonic Oscillator (SHO)

It is important to establish that the classical and quantum disparities found in this paper are not an artifact of an infinite gradient such as in the infinite square well for a free particle in a box; and that non-locality is also part of the QM SHO.

### 3.1 Classical Harmonic Oscillator

We begin with the classical harmonic oscillator so that we may compare with the corresponding expectation values for a quantum harmonic oscillator. Let us normalize the classical probability density $P$ which in classical mechanics (CM) is inversely proportional to the oscillating particle's velocity

$$1 = \int_{-A}^{A} \frac{b}{\pm w\left(A^2 - x^2\right)^{1/2}} dx \Rightarrow b = \frac{\pm w}{p}, \tag{3.1}$$

where b is the normalization constant, A is the classical amplitude, and the angular frequency $w = 2pf$. Therefore the normalized classical probability density is

$$bP = \frac{1}{p\left(A^2 - x^2\right)^{1/2}}. \tag{3.2}$$

The classical particle position expectation values are

$$\langle x \rangle_{CM} = \int_{-A}^{A} x \left[ \frac{1}{p\left(A^2 - x^2\right)^{1/2}} \right] dx = 0, \tag{3.3}$$

and all $\langle x^k \rangle_{CM}$ = 0 for odd values of k = 1, 3, 5, … because P is even and $x^k$ is odd for all odd *k*. For even values of *k*:

$$\langle x^2 \rangle_{CM} = \int_{-A}^{A} x^2 \left[ \frac{1}{p\left(A^2 - x^2\right)^{1/2}} \right] dx = \frac{A^2}{2}. \tag{3.4}$$

$$\langle x^4 \rangle_{CM} = \int_{-A}^{A} x^4 \left[ \frac{1}{p\left(A^2 - x^2\right)^{1/2}} \right] dx = \frac{3A^4}{8}. \tag{3.5}$$

$$\left\langle x^6 \right\rangle_{CM} = \int_{-A}^{A} x^6 \left[ \frac{1}{p\left(A^2 - x^2\right)^{1/2}} \right] dx = \frac{5A^6}{16}. \tag{3.6}$$

### 3.2 Quantum Harmonic Oscillator

The time independent Schrödinger equation for the SHO for a particle of mass m, oscillating with frequency $f$, and angular frequency $w = 2pf$, is:

$$\frac{-(h/2p)^2}{2m}\nabla^2 y + (2p^2 mf^2 x^2) y = Ey \tag{3.7}$$

The eigenfunction solution to Eq. (3.7) for the one-dimensional SHO is

$$y_n(x) = b_n e^{-\frac{x^2}{2}} H_n(x) = b_n e^{-\frac{a^2 x^2}{2}} H_n(ax) \ , \tag{3.8}$$

where $n$ = 0, 1, 2, 3,…, $x \equiv ax, a \equiv 2p\left[Mf/h\right]^{1/2} = \left[2pMw/h\right]^{1/2}$, and $H_n(x)$ is the Hermite polynomial of the $n$th degree in $x$:

$$H_n(x) = (-1)^n e^{x^2} \frac{d^n e^{-x^2}}{dx^n}. \tag{3.9}$$

In general, the normalization constant

$$b_n = \left[ \frac{a}{p^{1/2} 2^n n!} \right]^{1/2}. \tag{3.10}$$

We equate the quantum energy level solution to the classical energy

$$E_n = \left(n + \tfrac{1}{2}\right)hf = \left(n + \tfrac{1}{2}\right)h\left(w/2p\right) = \left(\tfrac{1}{2}\right)mw^2 A^2 \tag{3.11}$$

to help in the comparison of the classical and quantum position expectation values.

*3.2.1 Ground State n = 0 for Harmonic Oscillator*

Let us first examine the ground state expectation values $\langle x^k \rangle_{QM}$ since the variance with classical mechanics (CM) is expected to be the greatest here. The normalized eigenfunction for the ground state ($n$ = 0) is

$$y_0(x) = \frac{a^{1/2}}{p^{1/4}} e^{-\frac{a^2 x^2}{2}}. \tag{3.12}$$

In general, the expectation value of $\langle x^k \rangle_{QM0}$ is

$$\left\langle x^k \right\rangle_{QM\,0} = \int_{-\infty}^{\infty} y_0^* x^k y_0 dx = \int_{-\infty}^{\infty} x^k \left[ \frac{a^{1/2}}{p^{1/4}} e^{-\frac{a^2 x^2}{2}} \right]^2 dx. \qquad (3.13)$$

The expectation value of $\left\langle x^k \right\rangle_{QM} = 0$ for odd values of the index k = 1, 3, 5, …. because $y_0(x)$ is an even function and $x^k$ is odd. In general $\left\langle x^k \right\rangle_{QM} = \left\langle x^k \right\rangle_{CM} = 0$, and in particular $\left\langle x^k \right\rangle_{QM} = \left\langle x^k \right\rangle_{CM} = 0$ by symmetry in QM and CM.

$$\left\langle x \right\rangle_{QM\,0} = \int_{-\infty}^{\infty} x \left[ \frac{a^{1/2}}{p^{1/4}} e^{-\frac{a^2 x^2}{2}} \right]^2 dx = 0 = \left\langle x \right\rangle_{CM}. \qquad (3.14)$$

So let us focus on some even values of k.

$$\left\langle x^2 \right\rangle_{QM\,0} = \int_{-\infty}^{\infty} x^2 \left[ \frac{a^{1/2}}{p^{1/4}} e^{-\frac{a^2 x^2}{2}} \right]^2 dx = \frac{1}{2a^2} = \frac{A^2}{2} = \left\langle x^2 \right\rangle_{CM}. \text{ (\textbf{Accord with CM})} \qquad (3.15)$$

$$\left\langle x^4 \right\rangle_{QM\,0} = \int_{-\infty}^{\infty} x^4 \left[ \frac{a^{1/2}}{p^{1/4}} e^{-\frac{a^2 x^2}{2}} \right]^2 dx = \frac{3}{4a^4} = \frac{3A^4}{4} = 2\left\langle x^4 \right\rangle_{CM}. \qquad (3.16)$$

$$\left\langle x^6 \right\rangle_{QM\,0} = \int_{-\infty}^{\infty} x^6 \left[ \frac{a^{1/2}}{p^{1/4}} e^{-\frac{a^2 x^2}{2}} \right]^2 dx = \frac{15}{8a^6} = \frac{15A^6}{8} = 6\left\langle x^6 \right\rangle_{CM}. \qquad (3.17)$$

*3.2.2 First Excited State n = 1 for Harmonic Oscillator*

$$\left\langle x \right\rangle_{QM\,1} = \int_{-\infty}^{\infty} x \left[ \frac{a^{1/2}}{2^{1/2} p^{1/4}} (2ax) e^{-\frac{a^2 x^2}{2}} \right]^2 dx = 0 = \left\langle x \right\rangle_{CM}. \qquad (3.18)$$

$$\left\langle x^2 \right\rangle_{QM\,1} = \int_{-\infty}^{\infty} x^2 \left[ \frac{a^{1/2}}{2^{1/2} p^{1/4}} (2ax) e^{-\frac{a^2 x^2}{2}} \right]^2 dx = \frac{3}{2a^2} = \left\langle x^2 \right\rangle_{CM}. \text{ (\textbf{Accord with CM})} \qquad (3.19)$$

$$\left\langle x^4 \right\rangle_{QM\,1} = \int_{-\infty}^{\infty} x^4 \left[ \frac{a^{1/2}}{2^{1/2} p^{1/4}} (2ax) e^{-\frac{a^2 x^2}{2}} \right]^2 dx = \frac{15}{4a^4} = \frac{10}{9}\left\langle x^4 \right\rangle_{CM}. \qquad (3.20)$$

$$\left\langle x^6 \right\rangle_{QM\,1} = \int_{-\infty}^{\infty} x^6 \left[ \frac{a^{1/2}}{2^{1/2} p^{1/4}} (2ax) e^{-\frac{a^2 x^2}{2}} \right]^2 dx = \frac{105}{8a^6} = \frac{14}{9}\left\langle x^6 \right\rangle_{CM}. \qquad (3.21)$$

*3.2.3 Second Excited State n = 2 for Harmonic Oscillator*

$$\left\langle x \right\rangle_{QM\,2} = \int_{-\infty}^{\infty} x \left[ \frac{a^{1/2}}{2p^{1/4}2^{1/2}} \left(4a^2x^2 - 2\right) e^{-\frac{a^2x^2}{2}} \right]^2 dx = 0 = \left\langle x \right\rangle_{CM}. \tag{3.22}$$

$$\left\langle x^2 \right\rangle_{QM\,2} = \int_{-\infty}^{\infty} x^2 \left[ \frac{a^{1/2}}{2p^{1/4}2^{1/2}} \left(4a^2x^2 - 2\right) e^{-\frac{a^2x^2}{2}} \right]^2 dx = \frac{5}{2a^2} = \left\langle x^2 \right\rangle_{CM}. \text{ (**Accord with CM**)} \tag{3.23}$$

$$\left\langle x^4 \right\rangle_{QM2} = \int_{-\infty}^{\infty} x^4 \left[ \frac{a^{1/2}}{2p^{1/4}2^{1/2}} \left(4a^2x^2 - 2\right) e^{-\frac{a^2x^2}{2}} \right]^2 dx = \frac{39}{4a^4} = \frac{26}{25} \left\langle x^4 \right\rangle_{CM}. \tag{3.24}$$

$$\left\langle x^6 \right\rangle_{QM\,2} = \int_{-\infty}^{\infty} x^6 \left[ \frac{a^{1/2}}{2p^{1/4}2^{1/2}} \left(4a^2x^2 - 2\right) e^{-\frac{a^2x^2}{2}} \right]^2 dx = \frac{375}{8a^6} = \frac{6}{5} \left\langle x^6 \right\rangle_{CM}. \tag{3.25}$$

It is noteworthy that the quantum and classical second moments are equal, for all $n$, (although all the other even QM moments are greater due to barrier penetration). The Virial Theorem accounts for this unique SHO result because $\left\langle PE \right\rangle = \left\langle KE \right\rangle = \tfrac{1}{2} E_{total} = \tfrac{1}{2} k \left\langle x^2 \right\rangle$ in both CM and QM $\Rightarrow \left\langle x^2 \right\rangle_{QM} = \left\langle x^2 \right\rangle_{CM}$ , and as proven in detail next.

As shown earlier in Eq. (3.8) $y_n(x) = b_n e^{-\frac{x^2}{2}} H_n(x) = b_n e^{-\frac{a^2x^2}{2}} H_n(ax)$, where $x \equiv ax$, and $a \equiv 2p\left[Mf/h\right]^{1/2} = \left[2pMw/h\right]^{1/2}$.

$$\left\langle x^2 \right\rangle_{QM} = \left\langle \frac{x^2}{a^2} \right\rangle = \frac{1}{a^2} \int_{-\infty}^{\infty} y_n^* x^2 y_n \, dx = \frac{1}{a^2} \int_{-\infty}^{\infty} y_n^* \left[ \tfrac{1}{2}\left(x + \tfrac{d}{dx}\right) + \tfrac{1}{2}\left(x - \tfrac{d}{dx}\right) \right]^2 y_n \, dx$$

$$= \frac{1}{a^2} \int_{-\infty}^{\infty} y_n^* \left[ \tfrac{1}{4}\left(x + \tfrac{d}{dx}\right)^2 + \tfrac{1}{4}\left(x - \tfrac{d}{dx}\right)^2 + \tfrac{1}{4}\left(x + \tfrac{d}{dx}\right)\left(x - \tfrac{d}{dx}\right) + \tfrac{1}{4}\left(x - \tfrac{d}{dx}\right)\left(x + \tfrac{d}{dx}\right) \right] y_n \, dx \tag{3.26}$$

$$= \frac{1}{a^2} \int_{-\infty}^{\infty} y_n^* \left[ \tfrac{1}{4}\left(x + \tfrac{d}{dx}\right)^2 + \tfrac{1}{4}\left(x - \tfrac{d}{dx}\right)^2 + \tfrac{1}{2}\left( -\tfrac{d^2}{dx^2} + x^2 \right) \right] y_n \, dx$$

$$\left(x + \tfrac{d}{dx}\right)^2 y_n = 2\left[n(n-1)\right]^{1/2} y_{n-2}, \tag{3.27}$$

$$\left(x - \tfrac{d}{dx}\right)^2 y_n = 2\left[(n+1)(n+2)\right]^{1/2} y_{n+2}, \text{ and} \tag{3.28}$$

$$\int_{-\infty}^{\infty} y_n y_j \, dx = 0 \text{ for } n \neq j \tag{3.29}$$

because the Hermite polynomials are orthogonal, leaving only the 3rd term of the integrand in Eq. (3.26). Substituting, $x \equiv ax$:

$$
\begin{aligned}
\left\langle x^2 \right\rangle_{QM} &= \left(\frac{1}{a^2}\right)\int_{-\infty}^{\infty} y_n^* \left[\tfrac{1}{2}\left(-\frac{d^2}{a^2 dx^2} + a^2 x^2\right)\right] y_n \, dx \\
&= \int_{-\infty}^{\infty} y_n^* \left[-\left(\frac{h}{2(2pM)w}\right)^2 \frac{d^2}{dx^2} + x^2\right] y_n \, dx
\end{aligned}
\tag{3.30}
$$

$$
\begin{aligned}
&For the SHO: \left\langle Potential Energy \right\rangle_{QM} \equiv \left\langle PE \right\rangle_{QM} = \tfrac{1}{2} M w^2 \left\langle x^2 \right\rangle \\
&= \int_{-\infty}^{\infty} y_n^* \left[-\left(\frac{h^2}{2(4p^2 M)}\right)\frac{d^2}{dx^2} + \tfrac{1}{2} M w^2 x^2\right] y_n \, dx = \tfrac{1}{2}\int_{-\infty}^{\infty} y_n^* E_n y_n \, dx = \tfrac{1}{2} E_n = \tfrac{1}{2}\left(n + \tfrac{1}{2}\right)\frac{h}{2p} w
\end{aligned}
\tag{3.31}
$$

Since $\left\langle PE \right\rangle_{QM} + \left\langle KE \right\rangle_{QM} = E_n$, Eq. (3.31) $\Rightarrow \left\langle PE \right\rangle_{QM} = \left\langle KE \right\rangle_{QM} = \tfrac{1}{2} E_n$. Classically:

$$
\tfrac{1}{2} M w^2 \left\langle x^2 \right\rangle_{CM} = \left\langle PE \right\rangle_{CM} = \left\langle KE \right\rangle_{CM} = \tfrac{1}{2} E = \tfrac{1}{2} E_n . \tag{3.32}
$$

Therefore $\left\langle x^2 \right\rangle_{QM} = \left\langle x^2 \right\rangle_{CM}$ for all states of the SHO.

### 3.3 Comparison of Quantum and Classical Harmonic Oscillator

We now compare the quantum and classical harmonic oscillator position expectation values based upon Eqs. (3.4) to (3.6), and (3.14) to (3.32). It is noteworthy that $\left\langle x^2 \right\rangle_{CM} = \left\langle x^2 \right\rangle_{QM}$, although all higher order position even moments are not equal; and of course $\left\langle x^k \right\rangle_{QM} = \left\langle x^k \right\rangle_{CM} = 0$ for all odd k = 1, 3, 5, …. The accord of $\left\langle x^2 \right\rangle_{CM} = \left\langle x^2 \right\rangle_{QM}$ prevails for all quantum numbers.

The higher order CM position even moments are significantly smaller than the higher order QM position even moments, and the disparity increases as the moments get larger. This can be attributed to penetration of the quantum wave function into the classically forbidden region for both even and odd $y_n(x)$ as $y^* y = \left| y^2 \right|$ is even and enters into the integration. This effect will diminish as one goes to higher quantum states, and should disappear as $n \to \infty$ for pure states. It is not clear that this will happen for wave packets [17].

The significance of the difference in the classical and quantum higher order position moments is that Newton's Second Law of Motion is violated because the wave function penetrates the classically forbidden regions so that the particle spends less time in the central region and more time in the region of the classical turning points than allowed by Newton’s Second Law. Next let us look at the opposite case where a particle spends more time in the central region because the wave function terminates at the boundary rather than penetrating it.

## 4 Free Particle In A Box

The square well is an archetype problem of QM. It is used as a model for a number of significant physical systems such as free electrons in a metal, long molecule, the Wigner box, etc.

### 4.1 Quantum Case for Particle in a Box

The Schrödinger non-relativistic wave equation is:

$$\frac{-(h/2p)^2}{2m}\nabla^2 y + Vy = i(h/2p)\frac{\P}{\P t}y \,, \tag{4.1}$$

where $y$ is the wave function of a particle of mass m, with potential energy V. In the case of constant V, we can set V = 0 as only differences in V are physically significant. A solution of Eq. (4.1) for the one-dimensional motion of a free particle of nth state kinetic energy $E_n$ is:

$$y = b_n e^{i2px/l} e^{-i2pE_n t/h} = b_n e^{i2p\left(\frac{x}{l}-\frac{w}{2p}t\right)}, \tag{4.2}$$

where the wave function $y$ travels along the positive x axis with wavelength $l$, angular frequency $w$, and phase velocity $v = lw/2p$.

We shall be interested in the time independent solutions. The following forms are equivalent:

$$\begin{aligned} y_n &= b_n e^{i2px/l} = b_n \cos(2px/l) + i\sin(2px/l) \\ &= b_n \sin(npx/2a - np/2) \end{aligned}, \quad n = 1, 2, 3, \ldots. \tag{4.3}$$

for a particle in an infinite square well potential with perfectly reflecting walls at x = -a, and x = +a, so that $\frac{n}{2} l = 2a$. The wall length $2a$ can be arbitrarily large, but needs to be finite so that the normalization coefficient is non-zero.

We normalize the wave functions to yield a total probability of 1 for finding the particle in the region -a to +a, and find

$$1 = \int_{-a}^{a} y^* y \, dx = \int_{-a}^{a} |y|^2 dx \Rightarrow b_n = \frac{1}{\sqrt{a}} \tag{4.4}$$

where the normalization is independent of n.

In general

$$\left\langle x^k \right\rangle_{QM} = \int_{-a}^{a} y^* x^k y\, dx = \int_{-a}^{a} x^k |y|^2 dx, \text{ for k = 1, 2, 3, ….} \tag{4.5}$$

Since $|y|^2$ is symmetric **here** for both $y_{ns}$ and $y_{nas}$, $x^k|y|^2$ is antisymmetric in the interval -a to +a, because $x^k$ is antisymmetric. Thus without having to do the integration we know that $<x^k> = 0$ for all odd k, and in particular $<x> = 0$ for the nth state. Let us find the expectation values $<x^k>$ where for k = 1, 2, 4, and 6 for the free particle in the nth state.

$$\left\langle x \right\rangle_{QM} = \int_{-a}^{a} y^* x y\, dx = \int_{-a}^{a} x |y|^2 dx = 0. \tag{4.6}$$

$$\left\langle x^2 \right\rangle_{QM} = \int_{-a}^{a} y^* x^2 y\, dx = \int_{-a}^{a} x^2 |y|^2 dx = a^2\left[\frac{1}{3} - \frac{2}{p^2 n^2}\right] = \frac{a^2}{3}\left[1 - \frac{6}{p^2 n^2}\right]. \tag{4.7}$$

$$\left\langle x^4 \right\rangle_{QM} = \int_{-a}^{a} y^* x^4 y\, dx = \frac{a^4}{5} - \frac{4a^2\left(p^2 n^2 a^2 - 6a^2\right)}{p^4 n^4} = \frac{a^4}{5}\left[1 - \frac{20}{p^2 n^2} + \frac{120}{p^4 n^4}\right]. \tag{4.8}$$

$$\left\langle x^6 \right\rangle_{QM} = \int_{-a}^{a} y^* x^6 y\, dx = \frac{a^6}{7}\left[1 - \frac{42}{p^2 n^2} + \frac{840}{p^4 n^4} - \frac{5040}{p^6 n^6}\right]. \tag{4.9}$$

We will compare these values with the corresponding classical values in Sec. 4.2.

## 4.2 Classical Case for Particle in a Box

The classical probability P is inversely proportional to the velocity whose magnitude is constant throughout the box (except at the walls). Therefore P is uniform for finding a classical free particle in the region -a to +a. Normalizing the classical probability,

$$1 = \int_{-a}^{a} bP\, dx = bP(2a) \Rightarrow bP = \frac{1}{2a}. \tag{4.10}$$

As for the quantum case, classically $<x^k> = 0$ for all odd k because P is an even function. The classical expectation values of $<x>$ and $<x^2>$ are

$$\left\langle x \right\rangle_{ClassicalMechanics} = \left\langle x \right\rangle_{CM} = \int_{-a}^{a} x bP\, dx = \int_{-a}^{a} \frac{x}{2a} dx = 0. \tag{4.11}$$

$$\langle x^2 \rangle_{CM} = \int_{-a}^{a} x^2 bP dx = \int_{-a}^{a} \frac{x^2}{2a} dx = \frac{a^2}{3}. \qquad (4.12)$$

$$\langle x^4 \rangle_{CM} = \int_{-a}^{a} x^4 bP dx = \int_{-a}^{a} \frac{x^4}{2a} dx = \frac{a^4}{5}. \qquad (4.13)$$

$$\langle x^6 \rangle_{CM} = \int_{-a}^{a} x^6 bP dx = \int_{-a}^{a} \frac{x^6}{2a} dx = \frac{a^6}{7}. \qquad (4.14)$$

**4.3 Comparing QM and CM Cases for Complete Interval Expectation Values**

$$\langle x \rangle_{QM} = 0 = \langle x \rangle_{CM}. \qquad (4.15)$$

$$\langle x^2 \rangle_{QM} = \left[1 - \frac{6}{p^2 n^2}\right] \langle x^2 \rangle_{CM}. \qquad (4.16)$$

$$\langle x^4 \rangle_{QM} = \left[1 - \frac{20}{p^2 n^2} + \frac{120}{p^4 n^4}\right] \langle x^4 \rangle_{CM}. \qquad (4.17)$$

$$\langle x^6 \rangle_{QM} = \left[1 - \frac{42}{p^2 n^2} + \frac{840}{p^4 n^4} - \frac{5040}{p^6 n^6}\right] \langle x^6 \rangle_{CM}. \qquad (4.18)$$

It is clear from the analysis that the expectation values of all the odd moments $\langle x^k \rangle$ (k = 1, 3, 5, …) are exactly equal to 0 for both QM and CM. As one might expect, for even moments the variance between QM and CM is largest for small n, and furthermore is larger the higher the moment. It is also clear from Eqs. (4.16) to (4.18) that the QM even position moments approach the CM values as n gets large.

The result $\langle x \rangle_{QM} = 0 = \langle x \rangle_{CM}$ means that in moving between the walls of a box, a particle spends an equal amount of time on either side of the box and hence the expectation value for finding it, is at the center of the box. However, the results disagree for higher order moments such as $\langle x^2 \rangle_{QM} = \left[1 - \frac{6}{p^2 n^2}\right] \langle x^2 \rangle_{CM}$ for a particle in a perfectly reflecting box of length 2a between walls. At low quantum number *n*, this is smaller than the classical value $\langle x^2 \rangle_{CM} = \frac{a^2}{3}$ of Eq. (4.12). So, for the full well-width expectation value, this implies that not only does the particle spend an equal time on either side of the origin, but that the particle spends more time near the center of the box independent of the length a. This is inconsistent with the results for either full or partial range normalization. Since we can make the length *a* arbitrarily large, this effect is due to quantum mechanical non-locality of the presence of the walls making itself felt near the center of the box because it does not go away with large *a*. It is noteworthy that non-locality appears in such a fundamental case, as well as for the SHO.

This is a violation of Newton's First Law of Motion (NFLM) because the particle must slow down in the region of the origin even though there is a force on it only at the walls. The particle cannot both be going at a constant velocity between the walls, slow down near the center, and speed up again as it goes toward the opposite wall even if the walls are arbitrarily long. Therefore in this example, we have a quantum action on a particle even where there is no force. This is a simpler case than the Aharonov-Bohm [1], Aharonov-Casher [2], and similar effects, has many of the same elements, and may be even more intrinsic to QM. It is noteworthy that unlike such effects, it is independent of Planck's constant h; and there are no fields.

## 5 Quantum And Classical Periods

The object of this section is to relate QM phase and beat periods to CM periods.

### 5.1 Simple Harmonic Oscillator (QM Phase Period)

In general a wave packet representing a particle is given by a linear sum of the eigenfunctions for a given Hamiltonian

$$\Psi(x,t) = \sum_{n=1}^{\infty} b_n y_n(x) e^{-iwt} = \sum_{n=1}^{\infty} b_n y_n(x) e^{-i2pE_n t/h}, \tag{5.1}$$

because of the linearity of the Schrödinger equation. In particular for the simple harmonic oscillator, the energy eigenfunctions $y_n$ are given by Eq. (3.8) in terms of the Hermite polynomials. As we shall make a general argument here, it is not necessary to specify the particular eigenfunctions. We can see from Eq. (5.1) that the wave packet will complete $N$ full quantum mechanical phase periods, $Nt_{QM}$, when all the phase factors $e^{-i2pE_n t/h}$ are equal. Since $e^{-i2pE_n t/h} = \cos[2pE_n t/h] - i\sin[2pE_n t/h]$, this occurs when

$$2pE_n t/h = \frac{2pE_n N t_{QM}}{h} = 2pN + q, \tag{5.2}$$

where $q$ is the phase, and $N$ is an integer that may vary as a function of n. To satisfy Eq. (5.2), $q$ is either a constant, or only exceptional values of n may be used for the eigenfunctions that make up the wave packet. In the more general case $q$ = constant, so we may set $q = 0$ for convenience. Then, Eq. (5.2) implies

$$N t_{QM} = \frac{h}{E_n}[N] \Rightarrow t_{QM} = \frac{h}{E_n}, \tag{5.3}$$

where we are effectively considering one period with $N = 1$.

Thus from Eq. (5.3), quantum mechanically the phase period for the one-dimensional SHO wave packet is

$$t_{QM} = \frac{h}{E_n} = \frac{h}{(n+\frac{1}{2})(h/2p)w} = \frac{2p}{(n+\frac{1}{2})w}. \tag{5.4}$$

Classically the period is

$$t_{CM} = \frac{1}{f} = \frac{2p}{w}. \tag{5.5}$$

Taking the ratio of Eqs. (5.4) and (5.5):

$$\frac{t_{QM}}{t_{CM}} = \frac{2p}{(n+\frac{1}{2})w}\left[\frac{w}{2p}\right] = \frac{1}{(n+\frac{1}{2})} \xrightarrow[n\to\infty]{} 0. \tag{5.6}$$

For $n = 1$, $\frac{t_{QM}}{t_{CM}} = \frac{2}{3}$, and since the ratio decreases monotonically as $n$ increases, the two phase periods are never equal, and $t_{QM} < t_{CM}$.

**5.2 Free Particle in a Box (QM Phase Period)**

The QM energy levels peculiarly get further from the CM energy levels, for a free particle in a box. The QM energy dependence is

$$E = \frac{1}{2m}[p]^2 = \frac{1}{2m}\left[\frac{h}{l}\right]^2 = \frac{1}{2m}\left[\frac{h}{4a/n}\right]^2 = \frac{h^2}{2m}\left[\frac{n^2}{16a^2}\right] = E_1 n^2. \tag{5.7}$$

Because these energy levels go as $n^2$ they get further apart $\left[(n+1)^2 - n^2 = 2n+1\right]$ as $n$ increases unlike the classical continuum, and also unlike position expectation levels. This is also unlike the QM harmonic oscillator and most other potentials. However, it is not clear that this violates the Correspondence Principle if $h \to 0$ as $n \to \infty$, since the energy levels are proportional to $h^2n^2$. Otherwise energy states get further apart, while the position variance gets closer.

This peculiarity warrants a comparison of the classical and quantum periods. Classically the period for the one-dimensional motion of a particle of velocity *v* in a box of wall separation 2a is

$$t_{CM} = \frac{4a}{v} = \frac{4a}{\left[\frac{2E}{m}\right]^{1/2}} = 4a\left[\frac{m}{2E}\right]^{1/2}. \tag{5.8}$$

Now let us examine the quantum mechanical phase period. From the general argument by which Eq.(5.3) was derived for a wave packet:

$$t_{QM} = \frac{h}{E} = \frac{h}{E_n} = \frac{h}{E_1 n^2}. \tag{5.9}$$

Thus from Eqs. (5.2) and (5.3)

$$\frac{t_{QM}}{t_{CM}} = \frac{h}{E} \Big/ 4a\left[\frac{m}{2E}\right]^{1/2} = \frac{h}{E}\frac{1}{4a}\left[\frac{2E}{m}\right]^{1/2} = \frac{h}{2a\sqrt{2mE_1 n^2}} = \frac{h}{2an\sqrt{2mE_1}} \xrightarrow[n\to\infty]{} 0. \tag{5.10}$$

Note that $t_{QM} > t_{CM}$ for *n* = 1; $t_{QM} = t_{CM}$ for *n* = 2, and thereafter $t_{QM} < t_{CM}$. Except for the first 2 energy states, this trend is the same as the SHO for the phase $t_{QM}$.

**5.3 Quantum Beat Periods [Beat Period = (Beat Frequency)$^{-1}$]**

It is possible that the observable periods and hence the only periods relevant for the Correspondence Principle are associated with beats between the phases for adjoining energy states, i.e. $t_{QMb} = h/(E_{n+1} - E_n)$ in general, rather than the phase period $t_{QM} = h/E_n$ [cf. eqs. (5.4) and (5.9)] which may or may not be measurable. [This is analogous to the classical difference between phase velocity (which can be superluminal) and subluminal group velocity, where $v_p v_g = c^2$. The quantum beat frequency $w_{QMb}/2p = (E_{n+1} - E_n)/h = h(w/2p)[(n+1+1/2)-(n+1/2)]/h = (w/2p)$ is traditionally observed, e.g. atomic spectra.] For the Simple Harmonic Oscillator:

$$\left[\frac{t_{QMb}}{t_{CM}}\right]_{SHO} = \frac{2p/w_{QMb}}{2p/w} = \frac{2p/w}{2p/w} = 1 \text{ for all n. } (\textbf{Accord with CM}) \tag{5.11}$$

In this case for the Infinite Square Well:

$$\left[\frac{t_{QMb}}{t_{CM}}\right]_{ISW} = \frac{2n}{2n+1} \xrightarrow[n\to\infty]{} 1. \qquad (5.12)$$

For $n = 1$, $t_{CM} = 1.5 t_{QMb}$, and yet for the SHO $\left[t_{CM} = t_{QMb}\right]_{SHO}$ exactly for all n.

## 6 Findings In This Paper Are Similar to Established Effects

Although the Aharonov-Bohm [1] and Aharonov-Casher [2] effects are commonly thought to be explainable only by quantum mechanics (QM). there is also a classical interpretation. Even Berry's geometric phase [4] seems amenable to classical interpretation. It is not the purpose of this section to side with either the quintessential quantum, or classical explanations, but this will be by way of contrast, as the quantum-classical expectation value differences presented in this paper are not the result of electric or magnetic fields, or due to phase differences; and appear not to have classical explanations.

### 6.1 Aharonov-Bohm Effect

The question of which is more fundamental, force or energy is central to the foundations of physics, though it is somewhat rendered void in the Lagrangian or Hamiltonian formulations. In Newtonian classical mechanics (CM), force (*vis motrix* in Newton's Principia[14]), and kinetic energy (*vis viva* in Leibnitz' Acta erud.[12] ) are two of the foremost concepts. In QM, potential and kinetic energies are the primary concepts, with force hardly playing a role at all. It was not until 1959, some thirty-three years after the advent of QM that Aharonov and Bohm described gedanken electrostatic and magnetostatic cases in which physically measurable effects occur where presumably no forces act [1]. These are now known as the Aharonov-Bohm (A-B) effect.

In the magnetic case, an electron beam is sent around both sides of a long shielded solenoid or toroid so that the electron paths encounter no magnetic field and hence no magnetic force. Electrons do encounter a magnetic vector potential, which enters into the electron canonical momentum producing a phase shift of the electron wave function, and hence QM interference. If the electrons go through a double slit and screen apparatus the shielded magnetic field shifts the interference pattern periodically as a function of *h/e* in the shielded region, where *h* is Planck's constant and *e* is the electronic charge (in superconductors because of electron pairing, the magnetic flux quantum is *h/2e*).

This was confirmed experimentally and considered a triumph for QM. The A-B effect appears not to have been seriously challenged for forty-one years until 2000 when Boyer [5, 6] argued that the A-B effect can be understood completely classically. First he points out that there has been no real experimental confirmation of the A-B effect. The periodic phase shift of a two-slit interference pattern due to a shielded magnetic field has indeed been confirmed. However, no experiment has shown that there are no forces on the electrons, that the electrons do not accelerate, and that the electrons on the two sides of a solenoid (or toroid) are not relatively displaced.

Boyer [8, 9] then goes on to propose a classical mechanism. The electron induces a field in the conductor (shield or electromagnet) and this field acts back on the charged particle producing a force which speeds up the particle as it approaches and then slows the particle as it recedes, so that it time averages to 0. This sequence is reversed on the other side of the magnetic source producing interference. The displaced charge in the shield (or solenoid windings) affects the current in the solenoid, and hence the center-of-energy of the solenoid field.

### 6.2 Aharonov-Casher effect

In 1984 Aharonov-Casher [2] (A-C) proposed an analog of the A-B effect in which the electrons are replaced by neutral magnetic dipoles such as neutrons, and the shielded magnetic flux is replaced by a line charge. They claimed that the neutral magnetic dipole particles undergo a quantum phase shift and show an effect despite experiencing no classical force. The A-C effect has been confirmed experimentally, and although it is considered to be solely in the domain of QM, Boyer also proposed a classical interpretation of this effect.

In 1987 Boyer [10] argued that neutrons passing a line charge experience a classical electromagnetic force in the usual electric-current model for a magnetic dipole. This force will produce a relative lag between dipoles passing on opposite sides of the line charge, with the classical lag leading to a quantum phase shift as calculated by A-C. Boyer went on to predict that a consequence of his analysis is the breakdown of the interference pattern when the lag becomes comparable to the wave-packet coherence length.

In 1991, Mignani [13] showed that the A-C effect is a special case of geometrical phases, i.e. the standard Berry phase and the gauge-invariant Yang phase.

### 6.3 Berry's Geometric Phase

In 1984, the same year as the A-C effect, Berry [4] theoretically discovered that when an evolving quantum system returns to its original state, it has a memory of its motion in the geometric phase of its wavefunction. There are both quantum and classical examples of Berry's geometric phase (BGP), but as far as I know no one has yet challenged the QM case with a CM explanation. It is noteworthy that in 1992 Aharonov and Stern [3] did the QM analog of Boyer's [10] CM analysis, in examining BGP in terms of Lorentz-type and electric-type forces to show that BGP is analogous to the A-B effect.

## 7 Discussion

Although Quantum Mechanics (QM) is considered to be a theory that applies throughout the micro- and macro-cosmos, it has fared badly in the quantum gravity realm as discussed by Rabinowitz [15,16], and there is no extant theory after almost a century of effort. In the case of the macroscopic classical realm, it is generally believed that quantum expectation values should correspond to classical results in the limit of large quantum number n, or equivalently in the limit of Planck's constant $h \to 0$. Some processes thought to be purely and uniquely in the quantum realm like tunneling, can with proper modeling also exist in the classical realm as shown by Cohn and Rabinowitz [11].

Bohm has long contended that classical mechanics is not a special case of quantum mechanics [5, 6]. As shown by the analysis of the free particle in a box, and of the harmonic oscillator, the present paper makes an even stronger statement that the predictions of both Newton's First and Second Laws are violated in the quantum realm. So quantum mechanics is incompatible with them in that domain despite the fact that Newton's Second Law can be derived by QM [18]. Bohr's Correspondence Principle [7] formulated in 1928 argues that QM yields CM as the quantum number $n \to \infty$, though the energy levels for a particle in a box do not do so as shown in Sec. 5.2.

We can gain a new insight as to why no radiation is emitted in the ground state. From the perspective of beat frequencies, no radiation is emitted in the ground state because the wave function cannot represent observable oscillatory motion as there can be no difference in phase frequencies i.e. no beat frequency. This may be less tautological than saying "no radiation is emitted in the ground state because there is no lower state to go to."

## 8 Conclusion

The harmonic oscillator potential is archetypal in QM for blackbody radiation, specific heat of solids, etc.; and as an approximation to more difficult potentials, as a second order approximation to a Taylor series expansion near a stable equilibrium point. It is remarkable that the SHO is exactly solvable in all realms from CM to QM to quantum field theory. Thus it is a noteworthy accord in finding that $\langle x^2 \rangle_{CM} = \langle x^2 \rangle_{QM}$ and $t_{QMb} = t_{CM}$ (beat periods) exactly for the harmonic oscillator for all quantum numbers[17]; and probably for no other well. This occurs despite the fact that there is significant penetration of the wave function into the classically forbidden region. Despite these accords, the QM SHO exhibits non-locality, as does the FPB and all other potential wells.

Because of non-locality, the QM results contradict the very Hamiltonian from which they come, and a particle's velocity in an infinite square well varies, no matter how wide the box, nor how far the particle is from the walls (for negligible wall effects). If a particle's velocity, $v$, is constant, then its position probability distribution is obliged to be constant by isotropy and symmetry. If a particle's probability distribution is not constant, then its velocity is not constant since if $A \Rightarrow B$, then $not\,B \Rightarrow not\,A$. For a constant $v$, the Uncertainty Principle implies an indefinite position, but not a non-uniform position probability inside the well as this would violate isotropy and symmetry. It is not like a sound wave that must have nodes at a wall because the medium is clamped there. It is not like an electromagnetic field that is clamped at a conductor. The lack of uniformity implies that what we call the vacuum is a non-empty medium.

For well-width expectation values, the free particle in a box and the simple harmonic oscillator (SHO) are examined in detail to uncover classical and quantum disparities. Except for these simple cases, quantum mechanical solutions are exceedingly difficult and turbid. The results indicate that such discrepancies may be expected to be found commonly for a wide range of quantum phenomena. Quantum mechanics gives the illusion of obeying Newton's laws in the quantum realm because it starts with a Hamiltonian that incorporates Newton's law, and because QM can derive Newton's law (since it was formulated to do so). As shown in this paper, QM is incompatible with Newton's 1st and 2nd laws in the quantum domain, and this incompatibility appears to extend into the classical limit. Significant differences were found in this analysis for QM and CM expectation values. Since expectation values are supposed to correspond to possible classical measurements, one may be optimistic that these findings are amenable to experimental test. We should never underestimate the ingenuity of experimentalists, and the use of femtosecond lasers. There is the dilemma that the infinite well successive quantum energy levels get significantly further apart since $E_{n+1} - E_n \propto (n+1)^2 - n^2 = 2n + 1 \xrightarrow[n\to\infty]{} \infty$, in contrast to the classical continuum; as well as a significant difference in periods. For $n$ = 1, $t_{CM} = 1.5 t_{QMb}$, and yet for the SHO $[t_{CM} = t_{QMb}]_{SHO}$ exactly for all n.

This paper raises a question regarding the universality of QM, and whether apparent quantum self-inconsistency may be examined internally, or must be empirically ascertained. If there is an inherent lack of internal verifiability, this may either point to inconsistencies in quantum mechanics that should be fixed, or that nature is manifestly more non-classical than one would judge from the Hamiltonian used to obtain quantum solutions. The answer is not obvious.

## Acknowledgment

I wish to thank David Finkelstein, Art Cohn and Frank Rahn for helpful discussions; Brian Woodahl for checking some of the equations, and Michael Ibison for his interest.